\documentclass[preprint,eqsecnum,prd,nofootinbib]{revtex4}
\def\Journal#1#2#3#4{{#1} {\bf #2}, #3 (#4)}

\def\NPB{{\em Nucl. Phys.} B}
\def\PLB{{\em Phys. Lett.}  B}

\def\PRD{{\em Phys. Rev.} D}
\def\ZPC{{\em Z. Phys.} C}

\def\EPJC{{\em Eur. Phys. J.} C}

\def\lsim{\mathrel{\rlap{\lower4pt\hbox{\hskip1pt$\sim$}}
    \raise1pt\hbox{$<$}}}                
\def\gsim{\mathrel{\rlap{\lower4pt\hbox{\hskip1pt$\sim$}}
    \raise1pt\hbox{$>$}}}                

\def\d{\delta}

\def\p{\pi}

\def\G{\Gamma}

\def\bo{{\raise.15ex\hbox{\large$\Box$}}}         

\def\leftrightarrowfill{$\mathsurround=0pt \mathord\leftarrow \mkern-6mu
        \cleaders\hbox{$\mkern-2mu \mathord- \mkern-2mu$}\hfill
        \mkern-6mu \mathord\rightarrow$}       
\def\dvec#1{\vbox{\ialign{##\crcr
        \leftrightarrowfill\crcr\noalign{\kern-1pt\nointerlineskip}
        $\hfil\displaystyle{#1}\hfil$\crcr}}}           
\def\dilog#1{\,{\rm Li_2}#1}                        

\def\be{\begin{equation}}
\def\ee{\end{equation}}

\def\bex{\begin{displaymath}}
\def\eex{\end{displaymath}}

\def\bea{\begin{eqnarray}}
\def\eea{\end{eqnarray}}
\def\NO{\nonumber}

\def\beax{\begin{eqnarray*}}
\def\eeax{\end{eqnarray*}}

\usepackage{epsf}
\usepackage{graphicx}
\begin{document}
\preprint{FSU-HEP-100201}
\title{A Next-to-Leading-Order Study \\ of Dihadron Production}
\author{J.F. Owens}
\affiliation{Physics Department, Florida State University, 
Tallahassee, FL  32306}
\date{\today}
\begin{abstract}

The production of pairs of hadrons in hadronic collisions is studied 
using a next-to-leading-order Monte Carlo program based on the phase space 
slicing technique. Up-to-date fragmentation functions based on fits to 
LEP data are employed, together with several versions of current parton 
distribution functions. Good agreement is found with data for the dihadron 
mass distribution. A comparison is also made with data for the dihadron 
angular distribution. The scale dependence of the predictions and the 
dependence on the choices made for the fragmentation and parton distribution 
functions are also presented. The good agreement between theory and 
experiment is contrasted to the case for single $\pi^0$ production where 
significant deviations between theory and experiment have been observed.

\end{abstract}
\maketitle

\section{Introduction}
\label{sec:intro}

The many successes of QCD at describing large momentum transfer processes 
have helped establish it as the theory of the strong interactions. Indeed, 
largely due to this success, research concerning QCD has moved from testing 
the theory to testing the approximations used to obtain predictions from 
the theory. Even though the overall description of large momentum transfer 
processes appears to be satisfactory, there are still some systematic 
discrepancies between the theory and experiment. These include, 
for example, problems observed in direct photon production \cite{Au_photon} 
and single $\pi^0$ production \cite{Au_pizero}. One phenomenological approach 
has emphasized that the single 
particle production processes are sensitive to recoil corrections due to the 
emission of initial state radiation -- also known as $k_T$ smearing \cite{kT}. 
Another viewpoint \cite{Au_pizero} has stressed that, at least in the case 
of $\pi^0$ production, there 
may be problems with our knowledge of the fragmentation functions in the 
region where the momentum fraction, $z$, taken by the produced particle is 
large, since this is a region where the data to which these functions are 
fitted are limited. Also, it has been noted that this same high-$z$ region may 
require significant threshold resummation corrections. 

The production of hadron pairs relies on the same underlying dynamics as 
single particle production. Furthermore, the production of high-mass pairs 
relies on the same high-$z$ region of the fragmentation functions as does 
single particle production. Thus, if threshold corrections are important, 
or if the fragmentation functions are inadequately known, then one might 
expect to see comparable disagreement between the data for dihadron 
production and the predictions as one sees for single particle production.

It is the purpose of this paper to present a comparison between data for 
high-mass dihadron production and predictions based on QCD. These predictions 
have been obtained using a next-to-leading-order Monte Carlo based program 
which uses a variant of the phase space slicing technique \cite{bergmann,HO}. 
This allows 
the same kinematic cuts used in the extraction of the data to be imposed 
on the theoretical predictions. Another treatment of dihadron production at 
next-to-leading-order using Monte Carlo techniques may be found in 
\cite{CFG}. In addition, the related process of photon pair production, 
including photon fragmentation functions, may be found in \cite{BGPW}. The 
formalism in these two references has been used to investigate $\pi^0 \gamma 
{\rm \ and\ } \pi^0 \pi^0$ production at the LHC as reported in \cite{Binoth}.

The outline of the paper is as follows. In the next section a brief overview 
of the structure of the calculation is presented. Then, the predictions 
of the program are compared to data in Sec.~\ref{sec:comp}. A summary and 
some conclusions are presented in Sec.~\ref{sec:summary}.

\section{Next-to-leading-order Monte Carlo calculation}
\label{sec:calc}

The calculation described in this section is based on the two-cutoff phase 
space slicing technique describd in \cite{bergmann,HO}. The basic concept is 
to partition the three-body phase space into three regions using two cutoff 
parameters, $\delta_s$ and $\delta_c$. One region is where the 
$2 \rightarrow 3$ matrix elements have soft or collinear singularities, one 
contains hard-collinear singularites, and in the remainder the matrix 
elements are finite. In the soft and hard-collinear regions the matrix 
elements are approximated using the soft or leading pole approximations, 
respectively, and the variables describing the soft or collinear quanta can 
be integrated over analytically. The results have the same form as the 
lowest order $2 \rightarrow 2$ contributions, but depend explicitly on the 
cutoffs used to define the the soft and hard-collinear regions. Likewise, the 
remaining finite $2\rightarrow 3$ contributions depend on the cutoffs used to 
isolate the divergent regions. These two types of contributions are used to 
generate two-body and three-body weights which are added together at the 
histogramming stage. For infrared-safe observables the dependence on the 
cutoffs cancels, provided that sufficiently small values of the cutoffs are 
chosen. Specific examples of this procedure are given in \cite{HO}. 

The case of high-mass dihadron production is formally rather similar to that 
for single hadron production with the addition of another fragmentation 
function. The treatment presented here, therefore, follows closely the 
presentation given in Sec. III E. of Ref.~\cite{HO}. 

The input needed for this calculation includes 
the squared matrix elements for the $2 \rightarrow 3$ subprocesses 
and the results for the ${\cal O}(\alpha_s^3)$ one-loop contributions to the 
$2 \rightarrow 2$ subprocesses \cite{ks,es}. For the purpose of this 
example, the notation of \cite{ks} will be used, since much of the input 
needed can be found in the appendices of that paper. The partons are 
labelled as $A + B \rightarrow 1 + 2$ \ and $ A + B \rightarrow 1 + 2 +3$ \ 
for the $2 \rightarrow 2$ \ and $2 \rightarrow 3$ subprocesses, respectively. 
A flavor label $a_A$ is used to denote the flavor of parton $A$, and 
similarly for the other partons. 

The lowest-order contribution to the inclusive cross section for producing 
two hadrons $h_1 \mbox{\ and \ } h_2$ in a collision of hadrons of types 
$A$\ and $B$ can be written  
as 
\begin{eqnarray} 
d\sigma^B &=& \frac{1}{2 x_A x_B s} \sum_{a_A, a_B, a_1, a_2}
G_{a_A/A}(x_A) G_{a_B/B}(x_B) D_{h1/a_1}(z_1) D_{h_2/a_2}(z_2) dx_A\, dx_B\, 
dz_1\, dz_2\, \nonumber \\
& \times & \frac{(4 \pi \alpha_s)^2}{w(a_A) w(a_B)} \psi^{(4)}(\vec a,\vec p) 
\, d \Gamma_2 
\label{eqn:1pIBorn}
\end{eqnarray}
where $\vec a =\{a_A, a_B, a_1, a_2\}$\ and  $\vec p=\{p_A^{\mu},p_B^{\mu},
p_1^{\mu},p_2^{\mu}\}$ denote the sets of flavor indices and parton 
four-vectors, respectively. The factors appearing in the spin/color 
averaging are given by 
\[ w(a) = \left\{\begin{array}{ll}
2(1-\epsilon) V & \mbox{a=gluon} \\
2 N & \mbox{a=quark or antiquark} 
\end{array}
\right. \]   
with $N=3$\ and $V=N^2-1.$ The factor $d \Gamma_2$ is the differential 
two-body phase space element in $n$-dimensions,
\begin{equation}
d\G_2= \frac{d^{n-1}p_3}{2p^0_3(2\p)^{n-1}}
        \frac{d^{n-1}p_4}{2p^0_4(2\p)^{n-1}}
        (2\p)^n \d^n(p_1+p_2-p_3-p_4) \, .
\label{eqn:ps2}
\end{equation}
Eq.\ (\ref{eqn:1pIBorn}) 
gives the contribution where parton 1 fragments into the hadron $h_1$ and 
parton 2 fragments into $h_2$. Care must 
be taken to explicitly include in the sum over $\vec a$ those terms 
corresponding to the case where parton 2 fragments into $h_1$ and vice versa. 
For compactness, 
these terms will not be explicitly written. The squared matrix elements for 
the various subprocesses, denoted by $\psi^{(4)}(\vec a, \vec p)$, may be 
found in Ref.\ \cite{ks}.
 
Next, consider the one-loop virtual corrections to the $2 \rightarrow 2$ 
subprocesses. These take the form
\begin{eqnarray}
d\sigma^v &=& \frac{1}{2x_A x_B s} \sum_{a_A, a_B, a_1, a_2} 
G_{a_A/A}(x_A) G_{a_B/B}(x_B) D_{h_1/a_1}(z_1) D_{h_2/a_2}(z_2) dx_A\, dx_B\, 
dz_1\, dz_2\, \nonumber \\
& \times &\frac{(4\pi \alpha_s)^2}{w(a_A) w(a_B)}\left[ 
\frac{\alpha_s}{2 \pi} 
\left(\frac{4 \pi \mu^2_R}{2 p_A \cdot p_B}\right)^{\epsilon}
\frac{\Gamma(1-\epsilon)}{\Gamma(1-2\epsilon)}\right]
\psi^{(6)}(\vec a, \vec p) \, d\Gamma_2
\end{eqnarray}
where
\begin{eqnarray}
\psi^{(6)}(\vec a, \vec p) &=& \psi^{(4)}(\vec a, \vec p )\left[ -\frac{1}
{\epsilon^2}\sum_n C(a_n) - \frac{1}{\epsilon }\sum_n \gamma(a_n) 
\right] \NO \\
& + & \frac{1}{2 \epsilon}\sum_{m,n \atop m\ne n}
\ln \left( \frac{p_m\cdot p_n}{p_A \cdot p_B}\right)
\psi^{(4,c)}_{m,n}(\vec a, \vec p) \NO \\
& - & \frac{\pi^2}{6}\sum_n \psi^{(4)}(\vec a, \vec p) +\psi^{(6)}_{NS}
(\vec a, \vec p) + {\cal O}(\epsilon ).
\end{eqnarray}
This expression for $\psi^{(6)}$ differs slightly from Eq.\ (35) in Ref.\ 
\cite{ks} because we have chosen to extract a different $\epsilon$ dependent 
overall factor: a factor of $\Gamma(1+\epsilon ) \Gamma(1-\epsilon )
\approx 1 + \epsilon^2 \frac{\pi^2}{6}$ has 
been absorbed into the above expression for $\psi^{(6)}$. Furthermore, the 
arbitrary scale $Q_{ES}^2$ used in Ref.\ \cite{ks} has been chosen to be 
$2p_A \cdot p_B$. The expressions for 
the functions $\psi^{(4,c)}_{m,n}$ \ and $\psi^{(6)}_{NS}$ may be found in 
Appendix B of Ref.\ \cite{ks}. The quantities $C(a_n)$ \ and $\gamma(a_n)$\ 
are given by
\[ C(a) = \left\{\begin{array}{ll}
N = 3 & \mbox{a=gluon} \\
C_F = \frac{4}{3} & \mbox{a=quark or antiquark} 
\end{array}
\right. \]   
and
\[ \gamma(a) = \left\{\begin{array}{ll}
(11 N -2 n_f)/6 & \mbox{a=gluon} \\
3 C_F/2 & \mbox{a=quark or antiquark} 
\end{array}
\right. \]   
It will be convenient for subsequent expressions to adopt the following 
notation:
\begin{equation}
{\cal F}=\left(\frac{4\pi \mu^2_R}{2 p_A \cdot p_B}\right)^{\epsilon} 
\frac{\Gamma(1-\epsilon)}{\Gamma(1-2\epsilon)}.
\end{equation}
The one loop virtual contributions can now be written as 
\begin{eqnarray}
d\sigma^v &=& \frac{1}{2x_A x_B s} \sum_{a_A, a_B, a_1, a_2} 
G_{a_A/A}(x_A) G_{a_B/B}(x_B) D_{h_1/a_1}(z_1) D_{h_2/a_2}(z_2) dx_A\, dx_B\, 
dz_1\, dz_2  \nonumber \\
& \times &\frac{(4\pi \alpha_s)^2}{w(a_A) w(a_B)} {\cal F} 
\frac{\alpha_s}{2 \pi} 
\left(\frac{A_2^v}{\epsilon^2} + \frac{A_1^v}{\epsilon} + 
A_0^v\right) d\Gamma_2
\end{eqnarray}
where
\begin{eqnarray}
A_2^v &=& -\sum_n C(a_n) \psi^{(4)}(\vec a, \vec p)\\
A_1^v &=& -\sum_n \gamma(a_n) \psi^{(4)}(\vec a, \vec p) +\frac{1}{2}
\sum_{m,n \atop m\ne n}\ln \left(\frac{p_m \cdot p_n}{p_A \cdot p_B} \right)
\psi^{(4,c)}_{m,n}(\vec a, \vec p)\\
A_0^v &=& -\frac{\pi^2}{6}\sum_n C(a_n)\psi^{(4)}(\vec a,\vec p) + 
\psi^{(6)}_{NS}(\vec a,\vec p).
\end{eqnarray}

Next, the contributions from the $2 \rightarrow 3$ subprocesses in the 
limit where one of the final state gluons becomes soft are needed. The 
contributions of the $2 \rightarrow 3$ subprocesses may be written as 
\begin{eqnarray}
d\sigma^{2 \rightarrow 3} &=& \frac{1}{2 x_A x_B s} \frac{(4 \pi \alpha_s)^3}
{w(a_A) w(a_B)} {\cal F} \sum_{a_A, a_B, a_1, a_2, a_3} 
G_{a_A/A}(x_A) G_{a/B/B}(x_B) D_{h_1/a_1} (z_1) D_{h_2/a_2} (z_2) \nonumber \\
& \times & \Psi(a_A, a_B, a_1, a_2, a_3, p_A^{\mu},p_B^{\mu},p_1^{\mu},
p_2^{\mu},p_3^{\mu}) \, d\Gamma_3 \, dx_A \, dx_B \, dz_1 \, dz_2.
\label{eqn:1PI2to3} 
\end{eqnarray}
The expressions for the $2 \rightarrow 3$ squared matrix elements appearing 
in Eq.\ (\ref{eqn:1PI2to3}) may be found in Ref.\ \cite{es}.
As noted earlier for the two-body contributions, one must include in the sum 
all possible parton to hadron fragmentations. Here $d\Gamma_3$ is the 
three-body invariant phase space factor in $n-$dimensions:
\begin{equation}
d\G_3 = \frac{d^{n-1}p_3}{2p^0_3(2\p)^{n-1}}
        \frac{d^{n-1}p_4}{2p^0_4(2\p)^{n-1}}
        \frac{d^{n-1}p_5}{2p^0_5(2\p)^{n-1}}
        (2\p)^n \d^n(p_1+p_2-p_3-p_4-p_5) \, .
\label{eqn:ps3}
\end{equation}

Consider the case where the soft gluon is parton 3. In this limit, the 
function $\Psi$ may be expanded as:
\begin{equation}
\Psi(a_A, a_B, a_1, a_2, a_3, p_A^{\mu},p_B^{\mu},p_1^{\mu},p_2^{\mu},
p_3^{\mu}) \sim \sum_{m,n \atop m < n} \delta_{a_3,g} \frac{p_m \cdot p_n}
{p_m \cdot p_3 p_n \cdot p_3} \psi^{(4,c)}_{m,n}(a_A, a_B, a_1, a_2, 
p_A^{\mu}, p_B^{\mu}, p_1^{\mu}, p_2^{\mu}).
\end{equation}

Next, one must integrate over the soft region of phase space defined by 
$E_3 < \delta_s \sqrt{2 p_A \cdot p_B}/2$.  This is easily done using the 
integrals given in the appendix of Ref.~\cite{HO}. The resulting soft 
contribution may 
be written as 
\begin{eqnarray}
d\sigma^s &=& \frac{1}{2x_A x_B s} \sum_{a_A, a_B, a_1, a_2} 
G_{a_A/A}(x_A) G_{a_B/B}(x_B) D_{h_1/a_1}(z_1) D_{h_2/a_2} (z_2) dx_A\, 
dx_B\, dz_1\, dz_2\,  \nonumber \\
& \times &\frac{(4\pi \alpha_s)^2}{w(a_A) w(a_B)} {\cal F} 
\frac{\alpha_s}{2 \pi} 
\left(\frac{A_2^s}{\epsilon^2} + \frac{A_1^s}{\epsilon} + 
A_0^s\right) d\Gamma_2
\end{eqnarray}
where
\begin{eqnarray}
A_2^s &=& \sum_n C(a_n) \psi^{(4)}(\vec a, \vec p)\\
A_1^s &=& - 2 \ln \delta_s \sum_n C(a_n) \psi^{(4)}(\vec a, \vec p) 
-\frac{1}{2} \sum_{m,n \atop m\ne n}\ln \left(\frac{p_m \cdot p_n}
{p_A \cdot p_B} \right)\psi^{(4,c)}_{m,n}(\vec a, \vec p)\\ 
A_0^s &=& 2 \ln^2 \delta_s \sum_n C(a_n) \psi^{(4)}(\vec a, \vec p) \NO \\
&+& \left( \psi^{(4,c)}_{A,1}+\psi^{(4,c)}_{B,2} \right) 
\left[ \frac{1}{2} \ln^2 
\left( \frac{p_1 \cdot p_3}{p_A \cdot p_B} \right) + \dilog \left( 
\frac{p_2 \cdot p_3}{p_A \cdot p_B} \right) + 2 \ln \delta_s \ln \left( 
\frac{p_1 \cdot p_3}{p_A \cdot p_B} \right) \right] \NO \\
&+& \left( \psi^{(4,c)}_{A,2}+\psi^{(4,c)}_{B,1} \right) 
\left[ \frac{1}{2} \ln^2 
\left( \frac{p_2 \cdot p_3}{p_A \cdot p_B} \right) + \dilog \left( 
\frac{p_1 \cdot p_3}{p_A \cdot p_B} \right) + 2 \ln \delta_s \ln \left( 
\frac{p_2 \cdot p_3}{p_A \cdot p_B} \right) \right]. 
\end{eqnarray}

After the collinear singularities associated with the parton 
distribution functions and fragmentation function have been factorized 
and absorbed into the corresponding bare functions, there will be 
soft-collinear terms left over due to the mismatch between the integration 
limits of the collinear singularity terms and the factorization counterterms. 
Collecting together the various collinear 
terms, the result can be written as follows:
\begin{eqnarray}
d\sigma^{coll} &=& \frac{1}{2x_A x_B s} \sum_{a_A, a_B, a_1, a_2} 
G_{a_A/A}(x_A) G_{a_B/B}(x_B) D_{h_1/a_1}(z_1) D_{h_2/a_2} (z_2) dx_A\, 
dx_B\, dz_1\, dz_2\,  \nonumber \\
& \times &\frac{(4\pi \alpha_s)^2}{w(a_A) w(a_B)} {\cal F} 
\frac{\alpha_s}{2 \pi} 
\left(\frac{A_1^{coll}}{\epsilon} + A_0^{coll}\right)\, d\Gamma_2
\end{eqnarray}
where
\begin{eqnarray}
A_1^{coll} &=& \sum_n [2\ln \delta_s C(a_n) + \gamma(a_n)]\\
A_0^{coll} &=& \sum_{n=A,B}[2\ln \delta_s C(a_n) +\gamma(a_n)]\ln \left(\frac
{2p_A \cdot p_B}{\mu^2_f}\right) \nonumber \\ 
&+& \sum_{n=1,2}[2\ln \delta_s C(a_n) 
+\gamma(a_n)]\ln 
\left(\frac{2p_A \cdot p_B}{M^2_f}\right).
\end{eqnarray}
Here $\mu_f \mbox{\ and \ } M_f$ are the initial and final state factorization 
scales.

After the mass factorization has been performed, the bare parton distribution 
functions and fragmentation functions have been replaced by scale dependent 
$\overline {\mbox{MS}}$ functions. In addition, there are finite remainders 
involving functions $\widetilde G \mbox{\ and \ } \widetilde D$, expressions 
for which may be found in \cite{HO}:
\begin{eqnarray}
d\widetilde \sigma &=& \frac{1}{2 x_A x_B s} \sum_{a_A, a_B, a_1, a_2}
\frac{(4 \pi \alpha_s)^2}{w(a_A) w(a_B)} \frac {\alpha_s}{2 \pi} 
\psi^{(4)}(\vec a, \vec p) dx_A\, dx_B\, dz_1\, dz_2\, d\Gamma_2 
\nonumber \\
& \times & \left[\widetilde G_{a_A/A}(x_A,\mu^2_f) G_{a_B/B}(x_B,\mu^2_f) 
D_{h_1/a_1}(z_1,M^2_f) D_{h_2/a_2}(z_2,M^2_f) \right. \nonumber \\
& + & G_{a_A/A}(x_A,\mu^2_f) \widetilde G_{a_B/B}(x_B,\mu^2_f) 
D_{h_1/a_1}(z_1,M^2_f) D_{h_2/a_2}(z_2,M^2_f) \nonumber \\ 
& + & G_{a_A/A}(x_A,\mu^2_f) G_{a_B/B}(x_B,\mu^2_f) 
\widetilde D_{h_1/a_1}(z_1,M^2_f) D_{h_2/a_2}(z_2,M^2_f) \nonumber \\
& + & \left. G_{a_A/A}(x_A,\mu^2_f) G_{a_B/B}(x_B,\mu^2_f) 
D_{h_1/a_1}(z_1,M^2_f) \widetilde D_{h_2/a_2}(z_2,M^2_f)\right].
\end{eqnarray}

At this point, all of the singular terms have been isolated as poles in 
$\epsilon$ or have been factorized and absorbed into the bare parton 
distribution and fragmentation functions. The $\epsilon$ dependent pole 
terms all cancel amongst each other:
\begin{eqnarray}
& & A_2^v+A_2^s = 0 \\
& & A_1^v + A_1^s + A_1^{coll} = 0.
\end{eqnarray}
The finite two-body contribution is given by
\begin{eqnarray}
d\sigma^{2\rightarrow 2} &=& d \sigma^B + d\widetilde \sigma  \nonumber \\
&+& \frac{1}{x_A x_B s} \sum_{a_A, a_B, a_1, a_2} 
\frac {(4 \pi \alpha_s)^2}{w(a_A) w(a_B)} G_{a_A/A}(x_A,\mu^2_f) 
G_{a_B/B}(x_B,\mu^2_f) \nonumber \\ 
& \times & D_{h_1/a_1}(z_1,M^2_f) D_{h_2/a_2}(z_2,M^2_f)  
\frac {\alpha_s}{2 \pi} \left[ A_0^v + A_0^s + A_0^{coll}\right]
dx_A\, dx_B\, dz_1\, dz_2\, d\Gamma_2.
\end{eqnarray}
The three-body contribution, now evaluated in four dimensions, was given in 
Eq.\ (\ref{eqn:1PI2to3}) where now the soft and collinear regions of phase 
space are excluded.

The structure of the final result is two finite 
contributions, both of which depend on the soft and collinear cutoffs -- 
one explicitly and one through the boundaries imposed on the three-body 
phase space. However, when both contributions are added while calculating an 
observable quantity, all dependence on the cutoffs cancels when sufficiently 
small values of the cutoffs are used.

\section{Comparison to data}
\label{sec:comp}

Two sets of next-to-leading-order fragmentation functions have become 
available recently \cite{kkp,bfgw}. Both sets have been fit to high 
statistics data from $e^+e^-$ experiments. Accordingly, only charge symmetric 
combinations, e.g., $h^++h^-$, have been determined and the sets do not have 
fragmentation functions for individual charge states. Nevertheless, these 
sets can be used to generate predictions for experiments which measured 
either $\pi^0 \pi^0$ final states or symmetric combinations of charged 
hadrons. The NA-24 \cite{NA24}, CCOR \cite{CCOR}, and E-706 \cite{Begel} 
experiments each measured the production of $\pi^0$ pairs while the E-711 
\cite{E711} experiment measured the production of $\ h^+h^+, h^-h^-, {\rm\  
and} h^+h^-\ $ pairs. In the latter case, one can combine the tabulated 
results 
to give the cross section for producing the symmetric combination 
$(h^++h^-)+(h^++h^-)$.

\begin{figure}[t]
 \includegraphics[height=5in, angle=270]{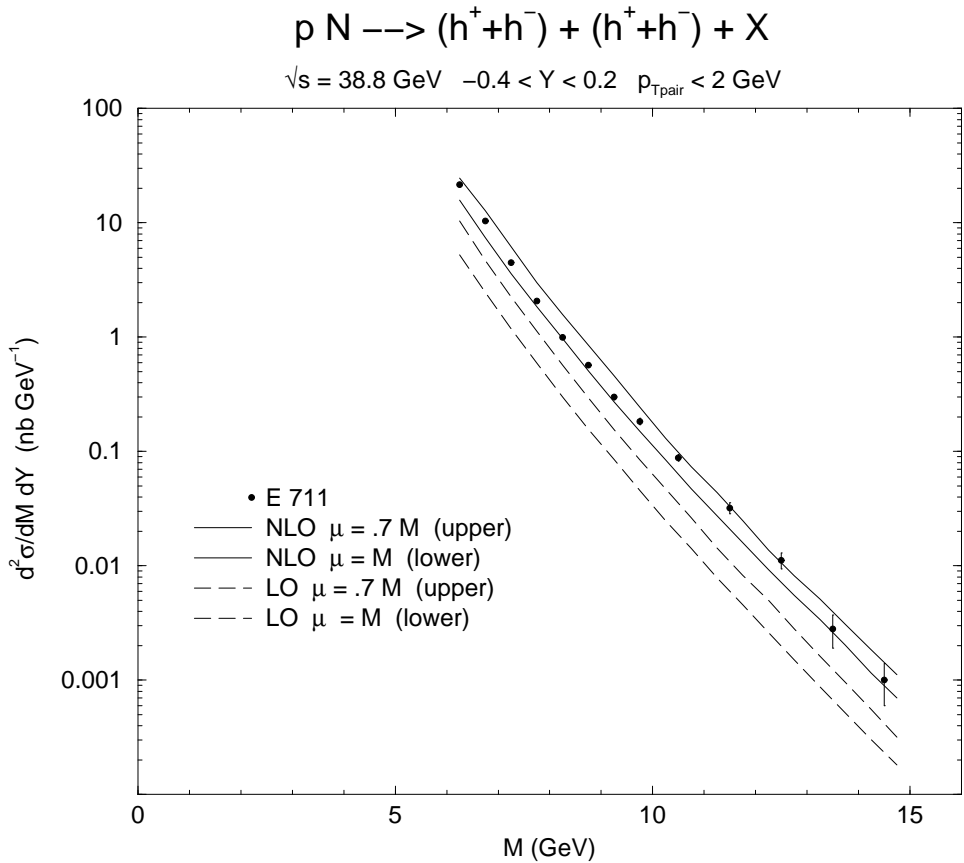}
 \caption{Comparison of the NLO(solid) and LO(dashed) results with data 
 from the E-711 experiment \cite{E711}.}
 \label{E711_fig1}
\end{figure}

In the following, unless otherwise stated, the theoretical results have been 
obtained using the CTEQ5M \cite{CTEQ5} parton distributions and the KKP 
\cite{kkp} fragmentation functions. For the calculation of the cross 
section at fixed values of the dihadron mass,$M$, 
the renormalization and factorization scales have been 
chosen to be proportional to $M$, as this is the only observed 
hadronic variable with the appropriate dimension.

\begin{figure}[!]
 \includegraphics[height=4in, angle=270]{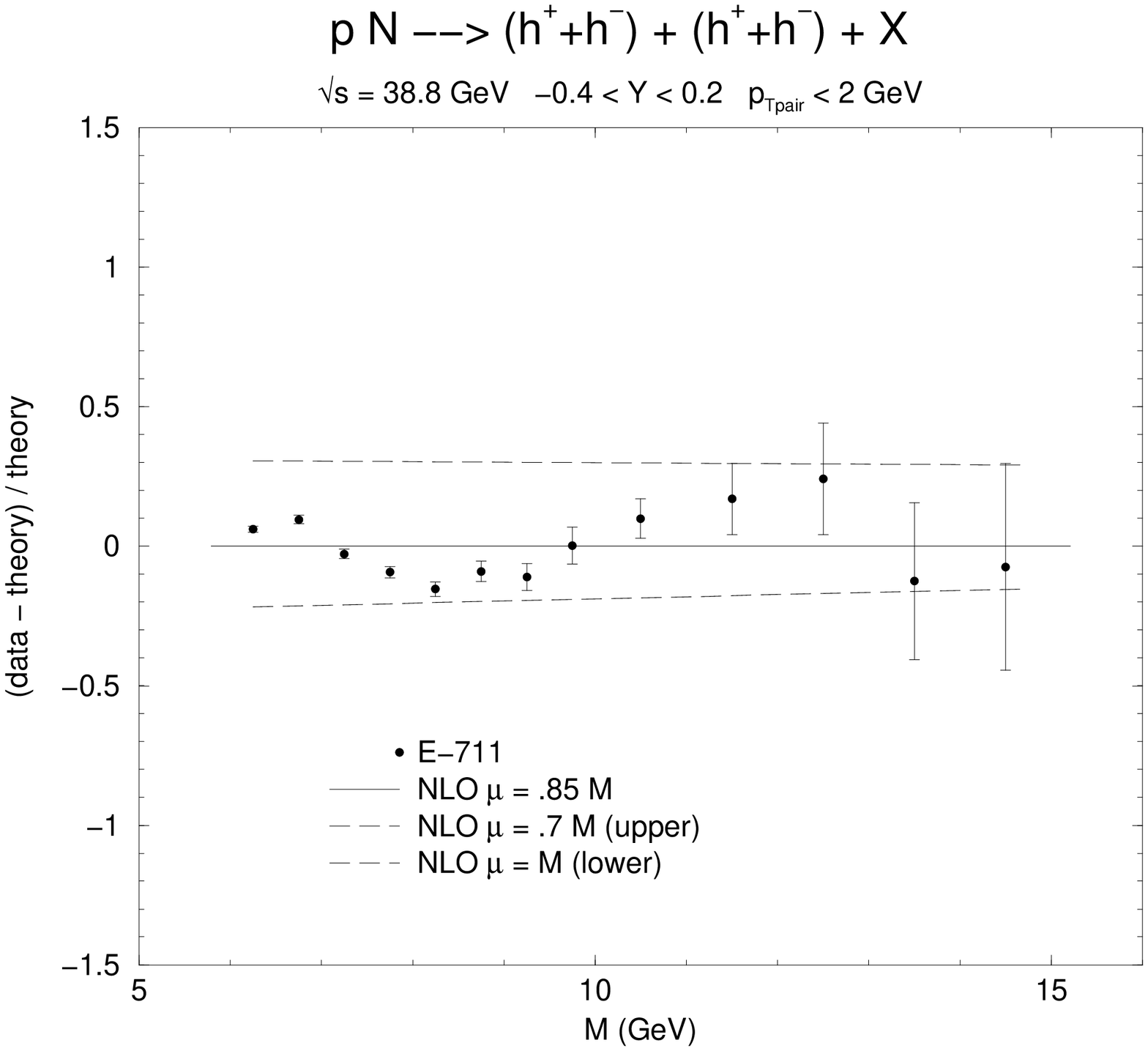}
 \caption{CTEQ5M results compared to the E-711 data for different scale 
choices. The renormalization and initial and final state factorization 
scales have been set equal to each other.}
 \label{E711_dmt}
\end{figure}

\begin{figure}[!]
 \includegraphics[height=4in, angle=270]{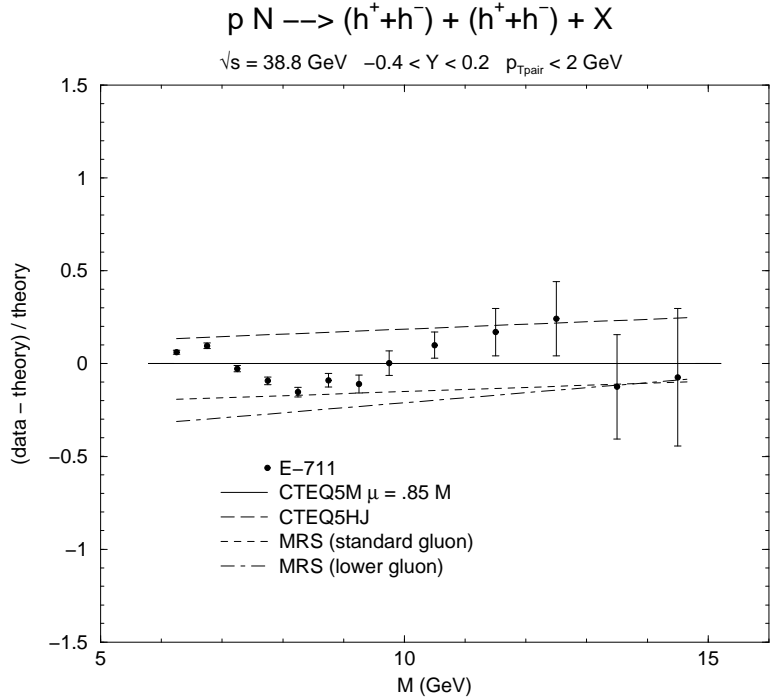}
 \caption{Comparison between the CTEQ5M predictions with those from the 
CTEQ5HJ \cite{CTEQ5}set and several MRST \cite{MRST} sets.}
 \label{E711_fig3}
\end{figure}

\begin{figure}[!]
 \includegraphics[height=4in, angle=270]{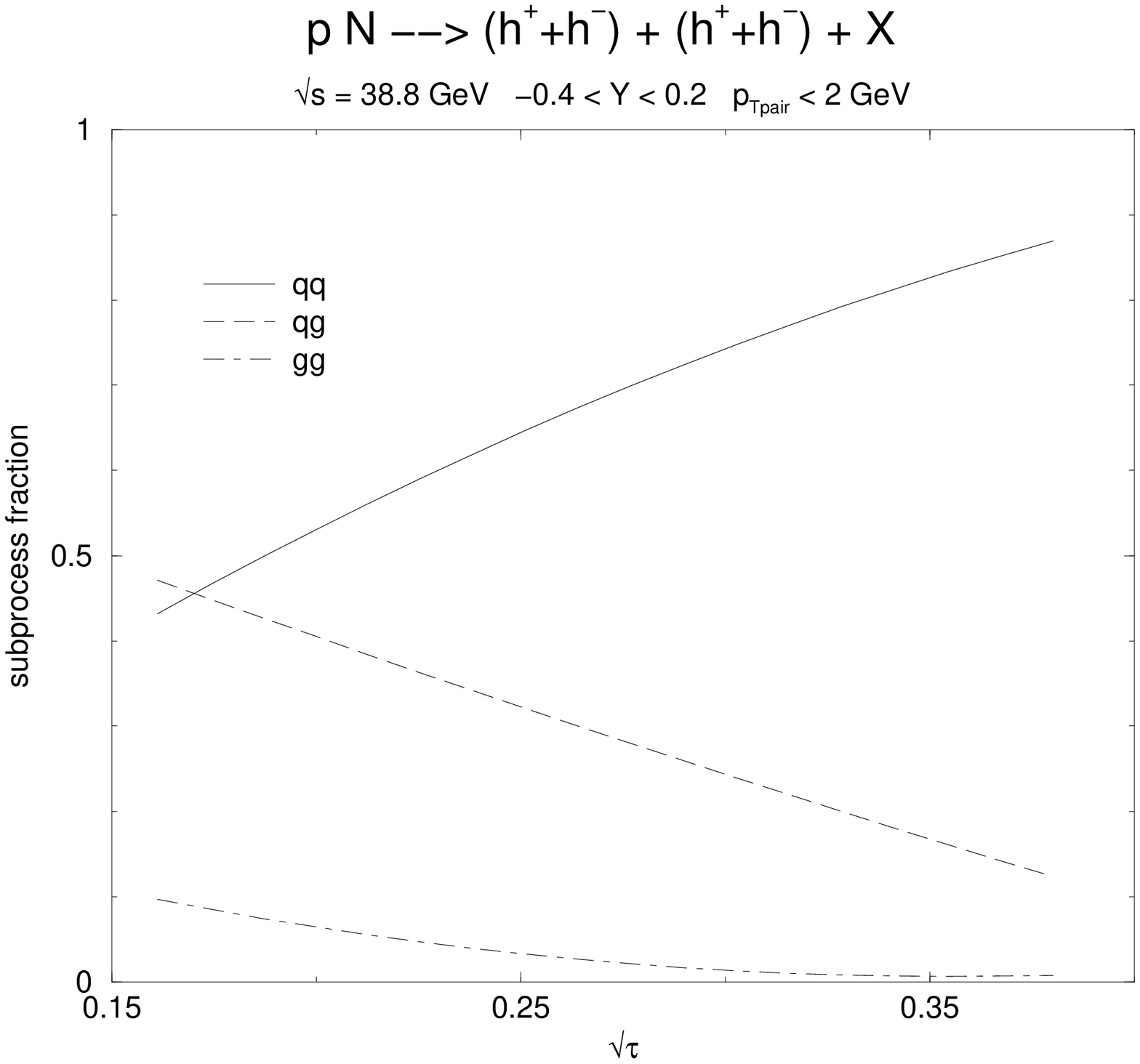}
 \caption{Relative contributions of the $qq, qg, {\rm \ and\ }gg$ 
subprocesses to the CTEQ5M results with $\mu=0.85 M$.}
 \label{subprocess}
\end{figure}

In FIG.~\ref{E711_fig1} the E-711 \cite{E711} data are shown with the 
leading-order (LO) 
and next-to-leading-order (NLO) theoretical results with two choices for the 
common factorization and 
renormalization scales. Cuts were applied to the rapidity of the pair, $Y$, 
the transverse momentum of the pair, $p_{Tpair}$, and $\cos \theta^*$, an 
estimate 
of the cosine of the scattering angle in the parton-parton center of mass 
frame. For these data, $\cos \theta^*$ was defined by first transforming to 
the frame where the momentum of the hadron pair had no longitudinal 
component. In general, the two hadrons will not be exactly back-to-back in 
this frame, due to their differing values of transverse momentum. The two 
values of the cosine of the angle between the hadron direction and the beam 
direction were averaged to obtain $\cos \theta^*$. The cuts used for the data 
shown in FIG.~\ref{E711_fig1} were $-0.4 < Y <  0.2, \ p_{Tpair}< 2 
{\rm \ GeV\ and\ } |\cos \theta^*|<0.25$. The NLO results can be seen to 
bracket 
the data while, for the scale choices shown, the LO results are 
significantly below the data. The large scale dependence evident at lowest 
order is due to the 
two powers of $\alpha_s$ in combination with the scale dependence of the four 
distribution and fragmentation functions. In the kinematic 
regime covered by the data, the distribution and fragmentation function 
momentum fractions are large, so that the functions decrease with increasing 
values of the scale. These scale dependences result in a significant 
decrease of the cross section with increasing scale, as shown in 
FIG.~\ref{E711_fig1}. The band covered by the corresponding NLO curves is 
narrower, although significant scale dependence remains. This is further 
examined in FIG.~\ref{E711_dmt} where the NLO results are compared to 
the E-711 data for three choices of scale. The format (data-theory)/theory 
is used in order to more clearly show the scale dependence. 

\begin{figure}[t]
 \includegraphics[height=4in, angle=270]{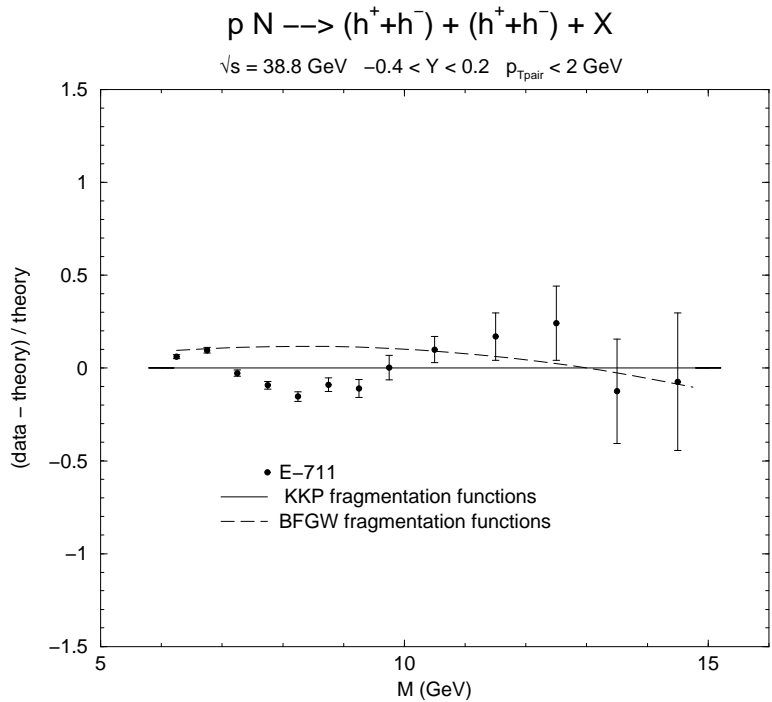}
 \caption{Comparison between the results obtained using the KKP \cite{kkp} 
and BFGW \cite{bfgw} fragmentation functions.}
 \label{E711_fig5}
\end{figure}

In FIG.~\ref{E711_fig3} the dependence of the results on variations in the 
choice of parton distributions is shown, relative to the CTEQ5M results. For 
each curve the scale has been set to $\mu = 0.85 M$. The 
CTEQ5HJ set \cite{CTEQ5} has a gluon distribution which has been enhanced in 
the high-$x$ region in order to better describe the high-$E_T$ jet data from 
the CDF and D\O\  Collaborations. Relative to the CTEQ5M distribution 
results,  
one can see an overall increase in the cross section, with the increase 
becoming larger towards the high mass end. Note that an increase in the 
scale from $0.85 M$ to $M$ would bring the CTEQ5HJ curve down to the level 
of the data. Also shown are the results for two of the MRST \cite{MRST} sets, 
one with the standard gluon and one with a reduced gluon. Although the 
two curves lie below the CTEQ5M results, a modest decrease in the scale choice 
would raise the curves to be in accord with the data. The results for a 
third MRST set with an increased gluon distribution are essentially 
identical with the CTEQ5M results.

In FIG.~\ref{subprocess} the relative contributions of the quark-quark, 
quark-gluon, and gluon-gluon subprocesses are shown versus $\sqrt\tau = 
M/\sqrt s$. 
This dimensionless variable is approximately the value of the parton momentum 
fraction which is probed in the production of the high mass hadron pair.
As expected, quark-quark scattering dominates at the upper end of the 
mass range covered by the data. Nevertheless, there is a significant 
contribution from quark-gluon scattering over much of the mass range. This 
is similar to the situation for high-$E_T$ jet production and, indeed, 
the results in FIG.~\ref{E711_fig3} do show some sensitivity to the choice of 
the parton distributions, e.g., CTEQ5M {\it vs.} CTEQ5HJ. Unfortunately, 
the scale dependence, even at NLO, is such as to preclude favoring one 
set over the other.

\begin{figure}
 \includegraphics[height=5in, angle=270]{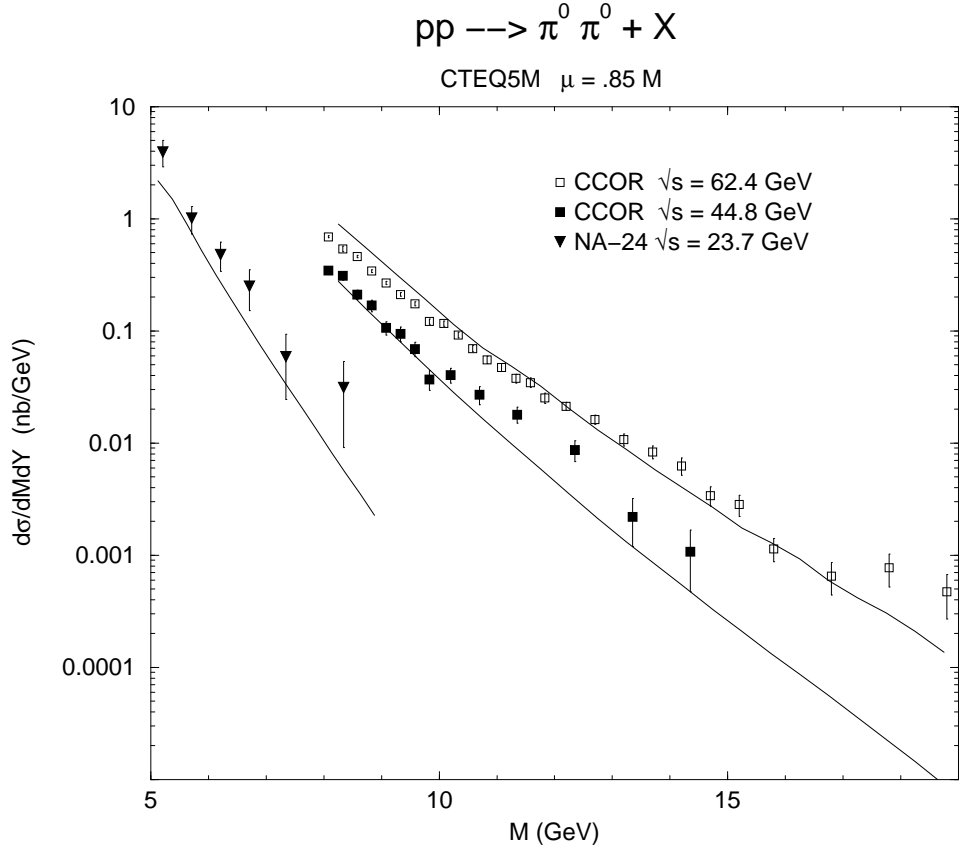}
 \caption{Comparison between the CTEQ5M results and data from the NA-24 
\cite{NA24} and CCOR \cite{CCOR} Collaborations.}
 \label{pi0pi0_mass}
\end{figure}

Next, in FIG.~\ref{E711_fig5} the dependence on the choice of the 
fragmentation functions is shown. The results from the two sets agree to 
within about 10\% across the mass range shown.

From the results shown thus far, several conclusions can be drawn. First, 
the NLO results give a very good description of the mass distribution 
for symmetric hadron pairs measured by the E-711 Collaboration. The 
variations observed due to different choices of the distribution and 
fragmentation functions are easily compensated for by changes in the 
renormalization and factorization scales. Nevertheless, extreme variations of 
these scales are not needed in order to describe the data.

\begin{figure}[!]
 \includegraphics[height=4in, angle=270]{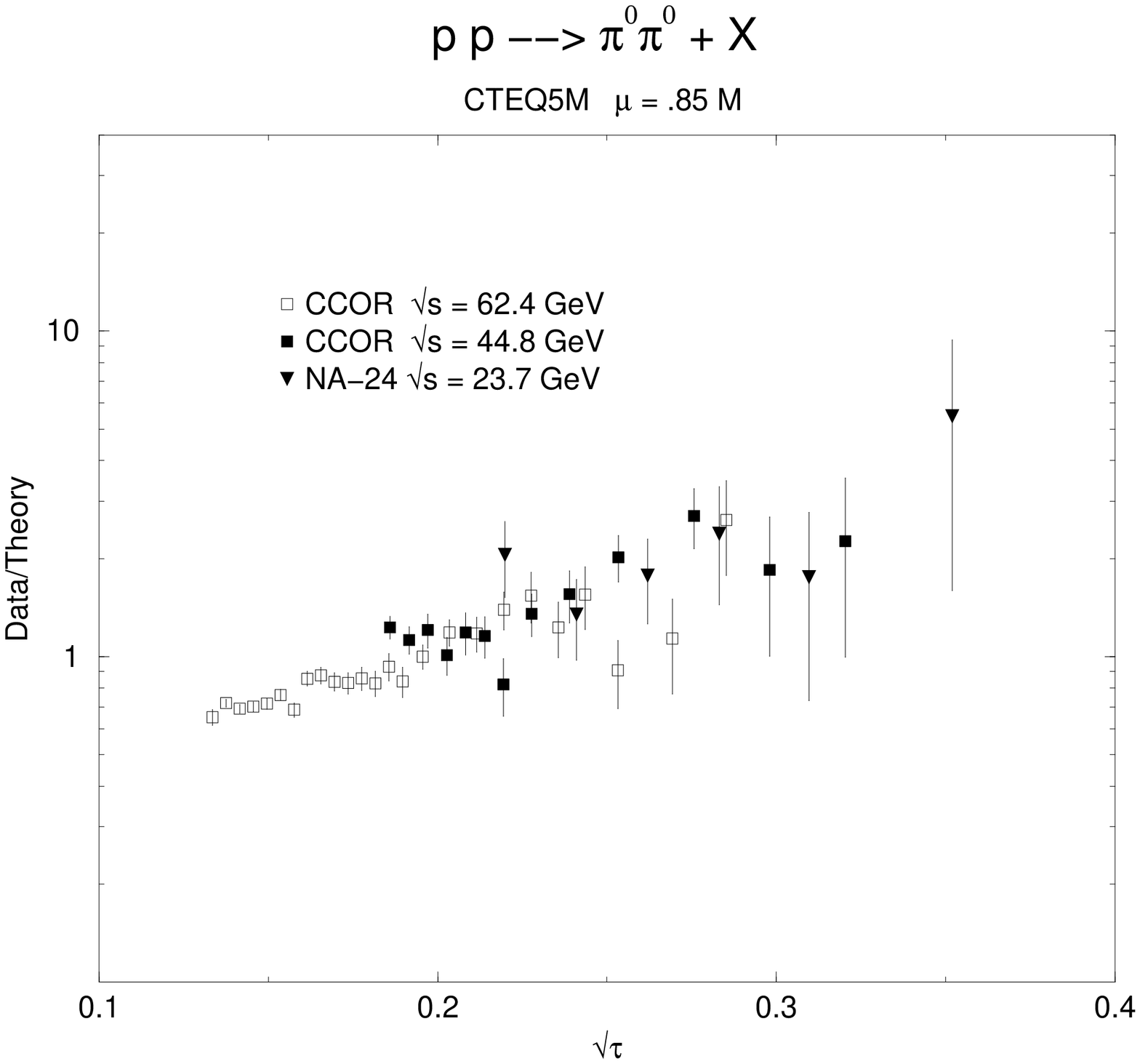}
 \caption{The same as for FIG.~\ref{pi0pi0_mass} presented in the 
 data/theory format.}
 \label{master_dot}
\end{figure}

\begin{figure}[!]
 \includegraphics[height=4in, angle=270]{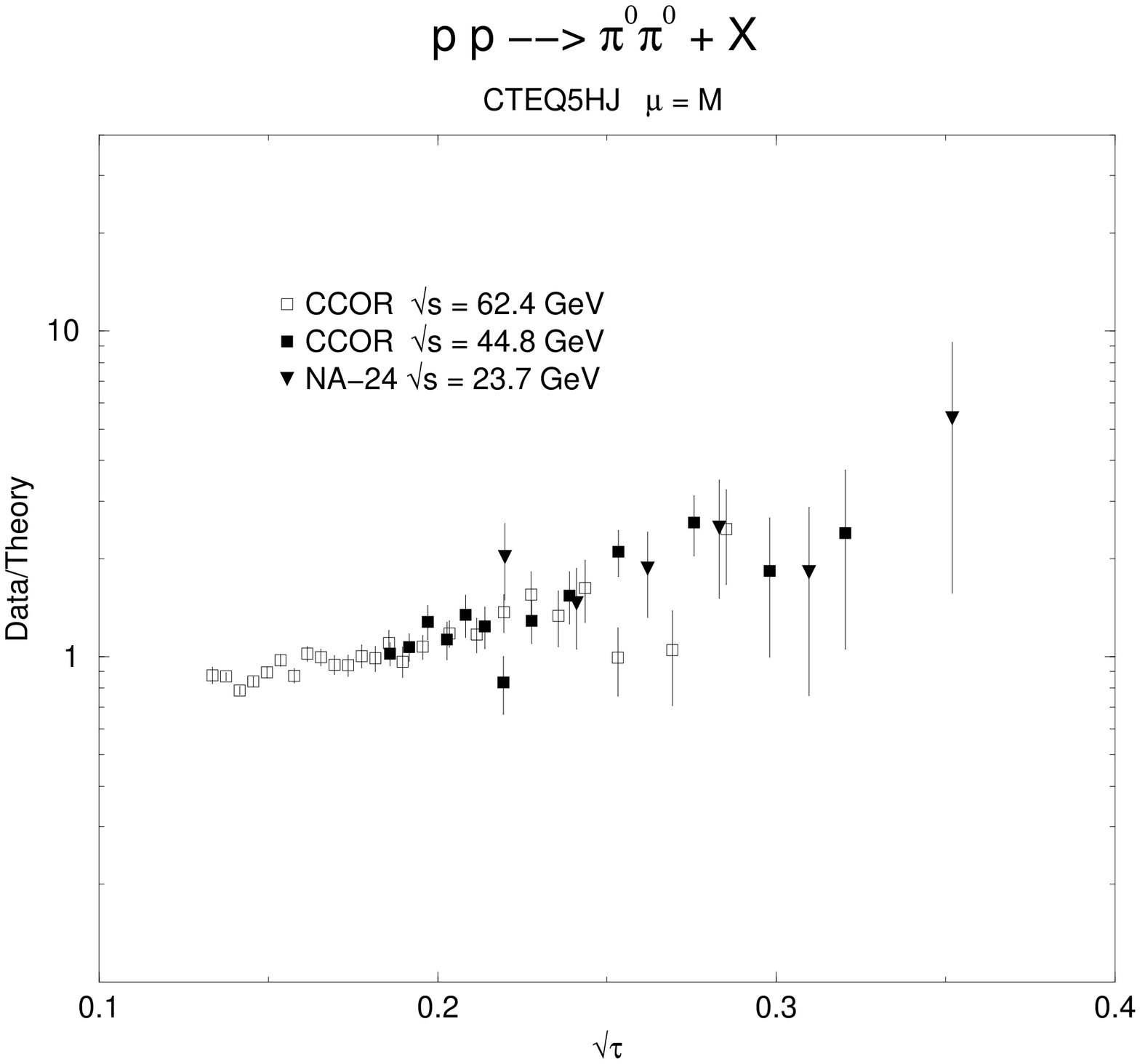}
 \caption{The same as for FIG.~\ref{master_dot} except using the CTEQ5HJ 
distributions.}
 \label{hj_dot}
\end{figure}

Next, consider the data for producing pairs of neutral pions.
In FIG.~\ref{pi0pi0_mass} the theoretical results are compared to data 
for the process $p p \rightarrow \pi^0 \pi^0 +X$ as measured by the 
NA-24 \cite{NA24} and CCOR \cite{CCOR} Collaborations. The cuts used for 
these data are $-0.35<Y<0.35, \ p_{Tpair} < 1, {\rm \ and\ } |\cos \theta^*| 
< 0.4$. The same scale choice of 
$\mu=0.85 M$ as used for the E-711 data gives a good description of the 
experimental results. Again, the 
CTEQ5M and KKP distribution and fragmentation functions have been used. 
The same results are shown in FIG.~\ref{master_dot} in a data/theory 
format versus $\sqrt\tau$. For the NA-24 data, the statistical and systematic 
errors have been added in quadrature. No discussion of errors was contained 
in the CCOR paper \cite{CCOR}. However, in an earlier publication on single 
$\pi^0$ production the CCOR Collaboration quotes an overall error of 25\% 
for both energies with an additional 5\% relative normalization error between 
the results for the two energies. Thus, the three data sets are seen to agree 
within the quoted errors. Furthermore, given these errors, the agreement with 
the theoretical results is acceptable, although deviations are apparent 
in FIGS.~\ref{pi0pi0_mass} and \ref{master_dot}. 

Noting the tendency for the 
data to lie below  the theory for small $\sqrt\tau $ and above it for larger 
$\sqrt\tau$, it is natural to ask whether the CTEQ5HJ distributions might 
yield better agreement with the data. These results are shown in 
FIG.~\ref{hj_dot} with a scale choice of $\mu=M$. Although the agreement 
is slightly better, the modified gluon distribution is unable to bring 
the theory and the data into complete agreement. Nevertheless, given the size 
of the errors, no definitive conclusion can be drawn.

\begin{figure}
 \includegraphics[height=5in, angle=270]{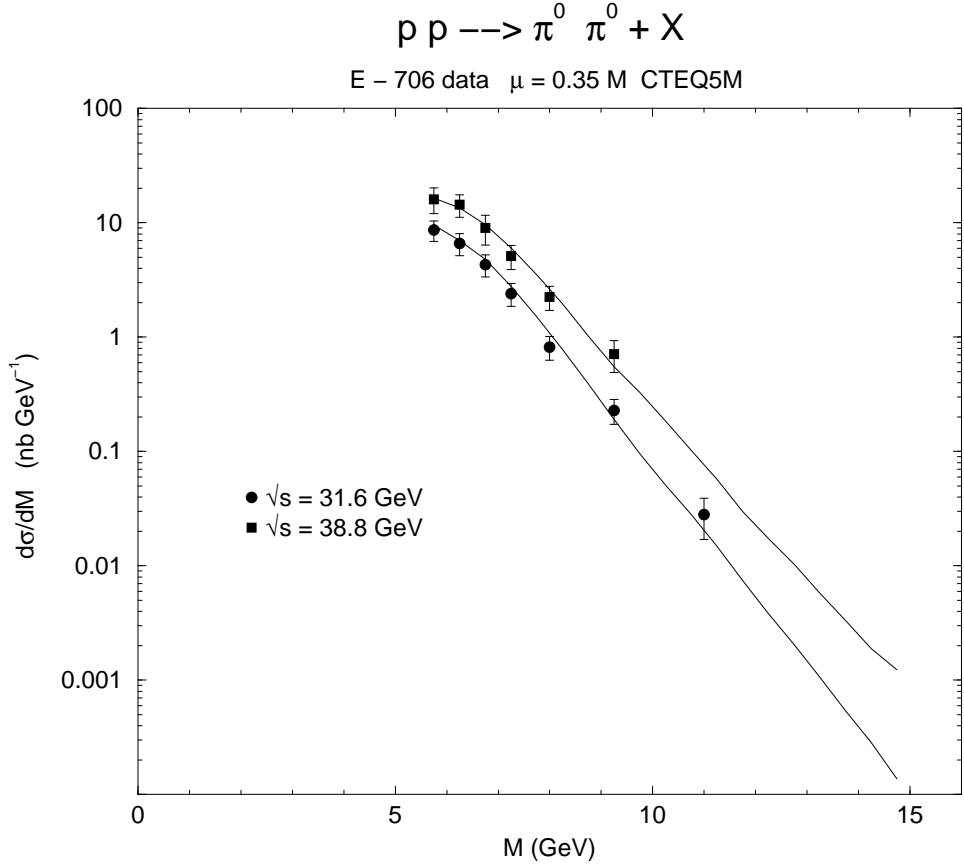}
 \caption{Comparison between the NLO results and data from the E-706 
Collaboration \cite{Begel}.}
 \label{E706}
\end{figure}

\begin{figure}
 \includegraphics[height=5in, angle=270]{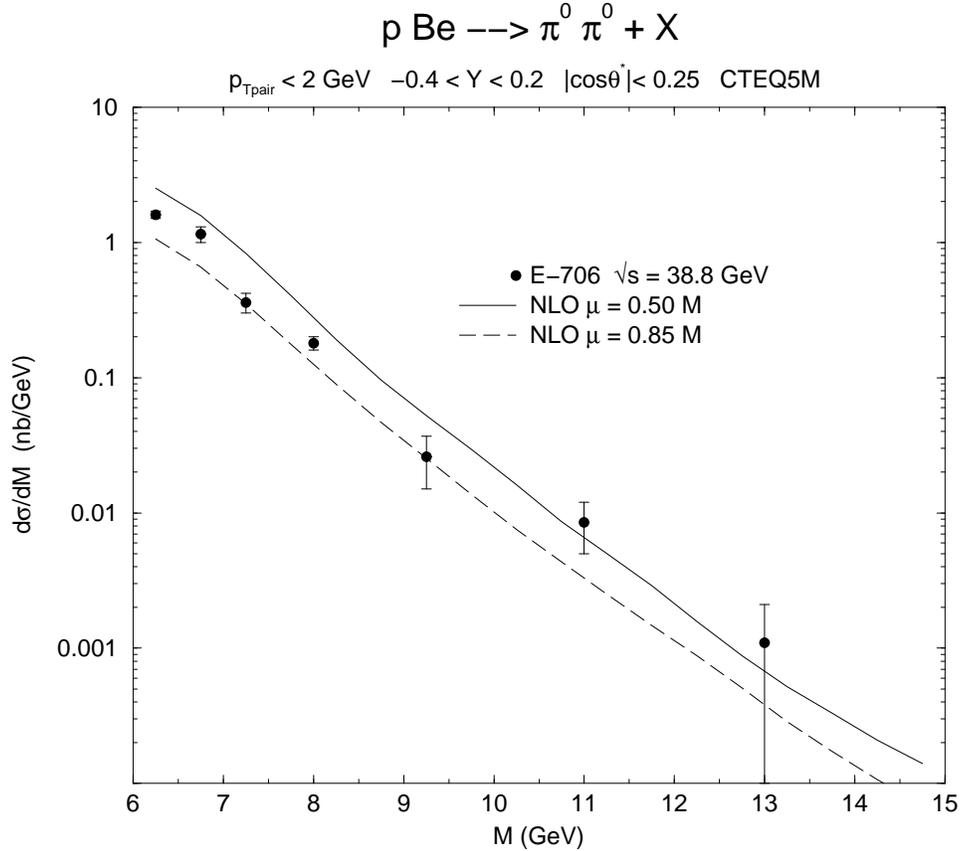}
 \caption{Comparison between the NLO results and data from the E-706 
Collaboration \cite{Begel}.}
 \label{E706_E711}
\end{figure}

In 
FIG.~\ref{E706} the NLO predictions are compared with data for dipion 
production as measured by the E-706 Collaboration \cite{Begel}. For these 
data the pions were separately required to satisfy $p_T > 2.5$ GeV and 
$-0.8 < y < 0.8 \ (-1.05 < y < 0.55)$ for the $\sqrt{s} = 31.6 \ (38.8)$ GeV 
data. The difference between the azimuthal angles for the two pions, 
$\Delta \phi$, was required to be greater than 
105 degrees. No cuts were placed on $\cos\theta^*$ or on $p_{Tpair}$. 
There is good agreement between the theory and the data when a scale choice 
of 
$\mu = 0.35 M$ is used. This value is significantly smaller than that used in 
the previous comparisons. Of course, the cuts used for the E-706 data set are 
far different than those used for the other sets. As a consistency check on 
their data the E-706 Collaboration \cite{Begel} also presented results 
obtained using 
cuts similar to those used for the E-711 data. These data are compared to 
the NLO predictions in FIG.~\ref{E706_E711} with $\mu=$ 0.50 $M$ and 0.85 $M$.
The two curves bracket the data, suggesting that E706 results are compatible 
with those of the other experiments when the same cuts are used. This suggests 
that the need for a much smaller scale when comparing with the E-706 data in 
FIG.~\ref{E706} is a problem due to the calculation not being able to 
properly reproduce the effects of the different sets of cuts. 

\begin{figure}[t]
\includegraphics[height=4in, angle=270]{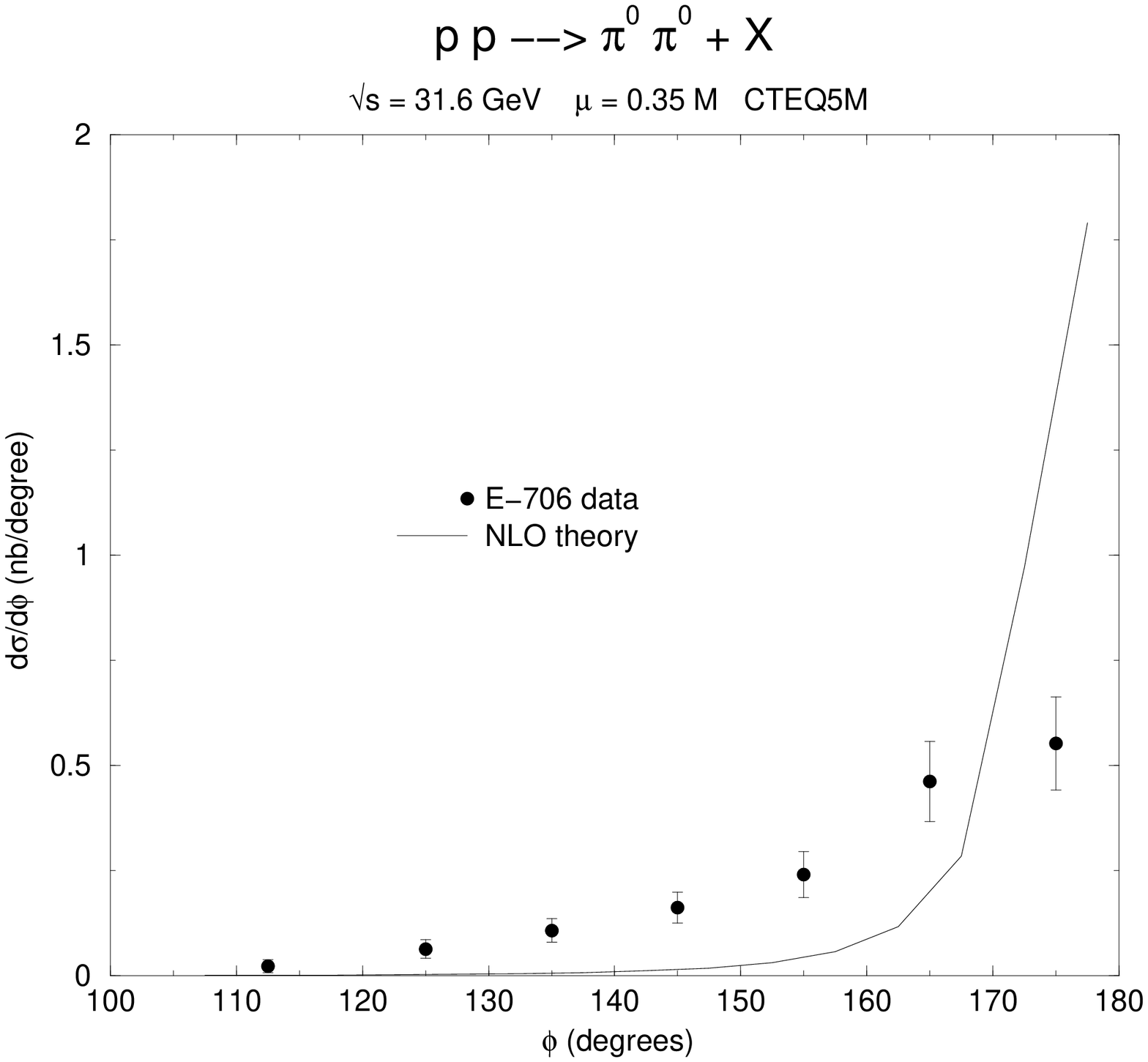}
\caption{Comparison of the NLO $\Delta \phi$ distribution with data from E-706 
\cite{Begel}.}
\label{E706_phi}
\end{figure}

\begin{figure}[!]
\includegraphics[height=4in, angle=270]{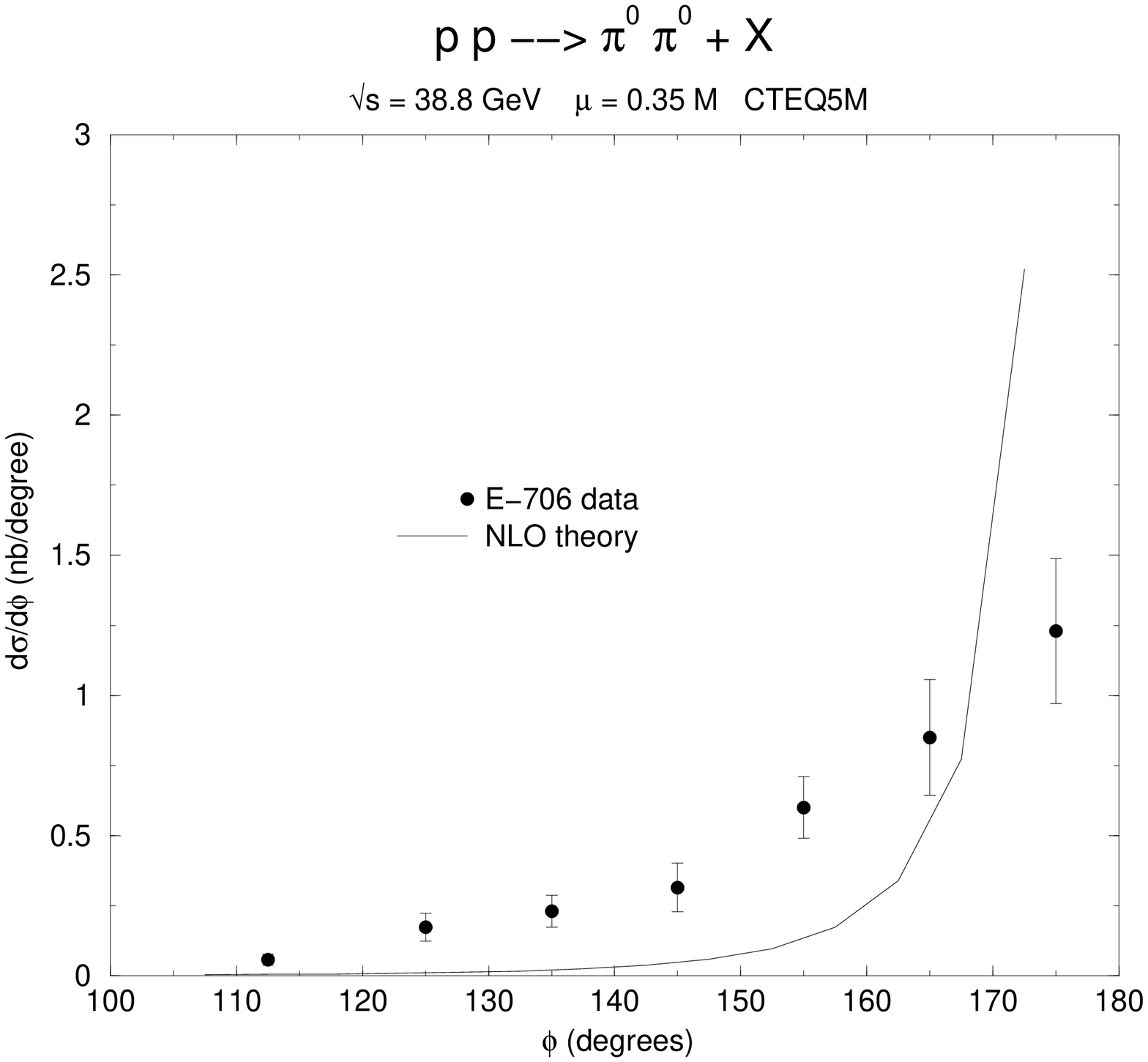}
\caption{Comparison of the NLO $\Delta \phi$ distribution with data from E-706 
\cite{Begel}.}
\label{E706_phi_800}
\end{figure}

\begin{figure}[t]
\includegraphics[height=4in, angle=270]{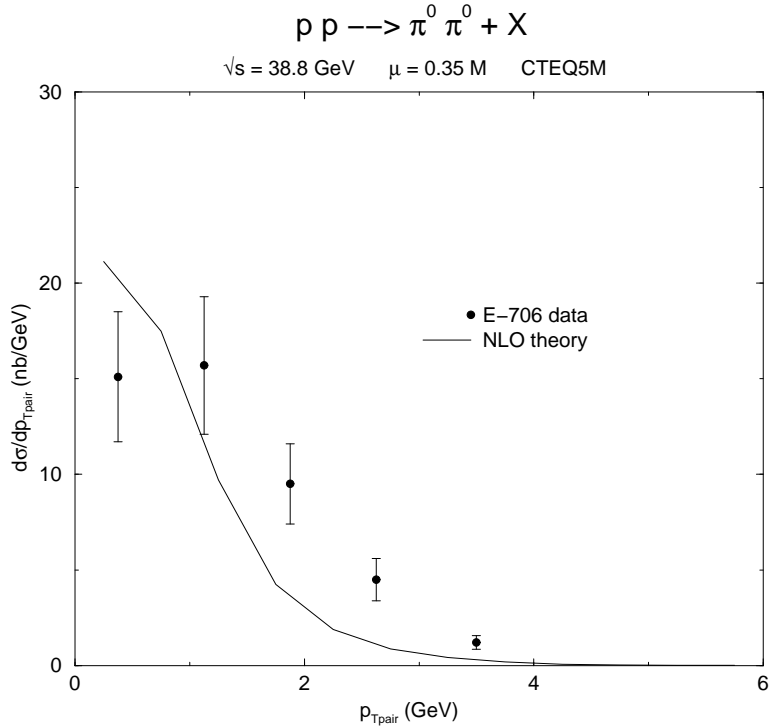}
\caption{Comparison of the NLO $p_{Tpair}$ distribution with data from E-706 
\cite{Begel}.}
\label{E706_qt}
\end{figure}

\begin{figure}[t]
 \includegraphics[height=5in, angle=270]{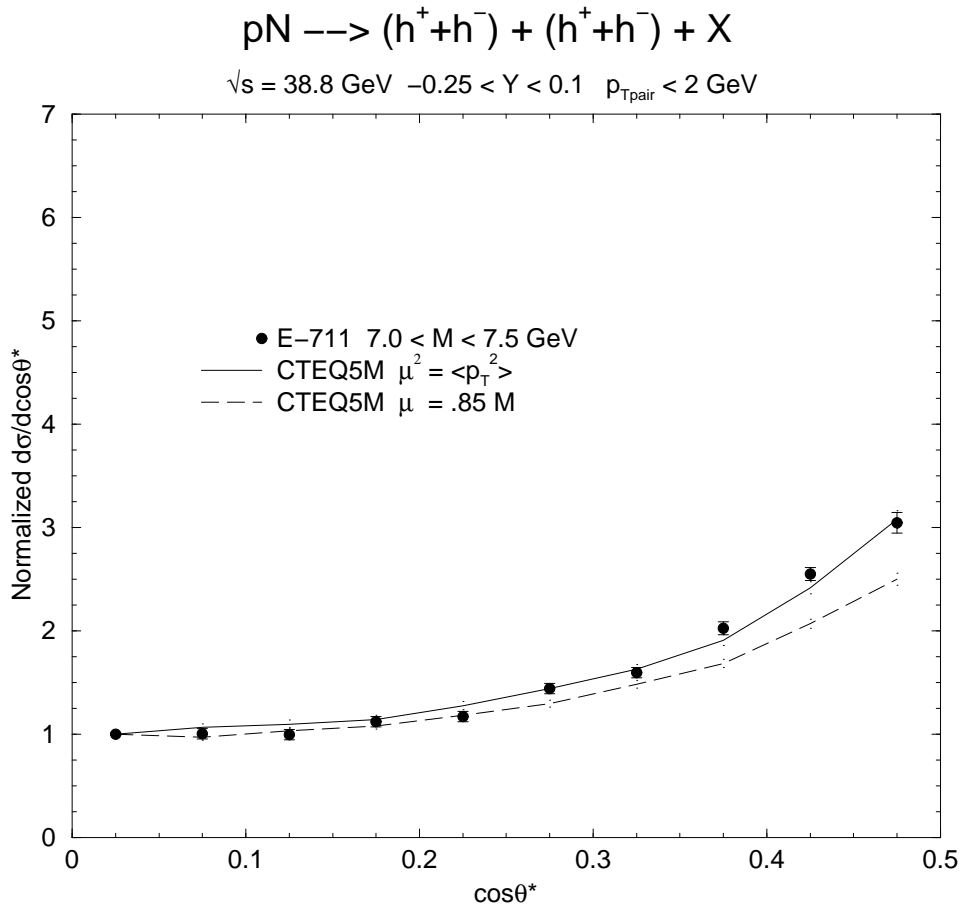}
 \caption{Comparison between the CTEQ5M results and the E-711 angular 
distribution data for $7 < M < 7.5$ GeV using two choices for the scale 
parameter.}
 \label{dcost4}
\end{figure}

\begin{figure}
 \includegraphics[height=5in, angle=270]{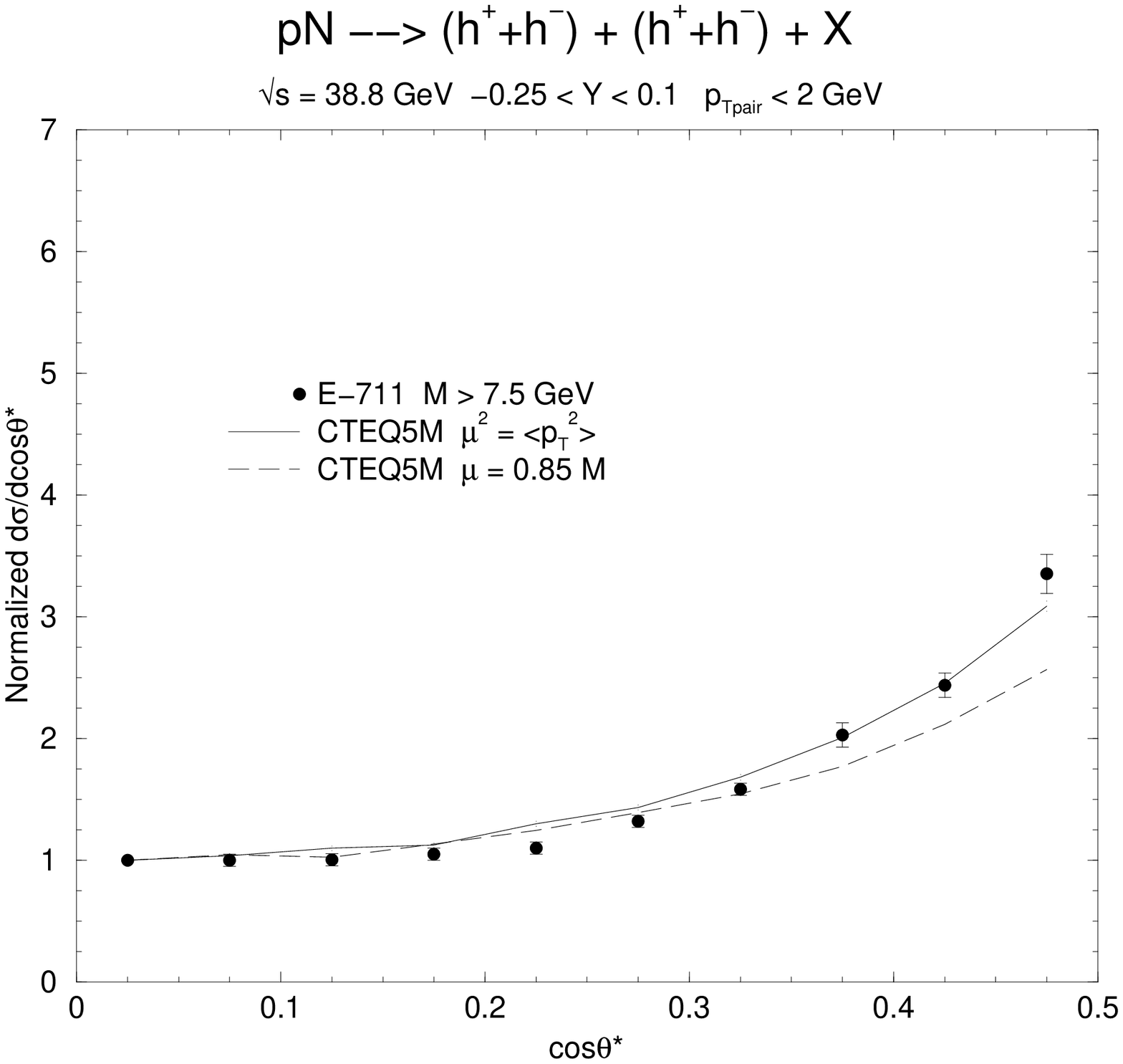}
 \caption{Same as for FIG.~\ref{dcost4} except for $M > 7.5$ GeV.}
 \label{dcost3}
\end{figure}

In order to investigate this situation further, consider the effects of a cut 
in $\Delta \phi$.
The signifigance of the $\Delta \phi$ cuts lies in the fact that the observed 
$\Delta \phi$ distributions are broader than those given by the NLO 
calculations, as can be seen in FIGS.~\ref{E706_phi} and \ref{E706_phi_800}. 
At lowest order, with collinear fragmentation and distribution 
functions, the theoretical predictions have the two hadrons being produced 
back-to-back, {\em i.e.,} with $\Delta \phi=180$ degrees. At 
next-to-leading-order, the $2 \rightarrow 3$ subprocesses allow for the 
$\Delta \phi$ distribution to develop a non-zero width. Nevertheless, it is 
still narrower than the experimental observations. The acceptance for the 
CCOR and E-711 experiments was limited to $\Delta \phi > 140$ degrees, and 
this cut was also placed on the NA-24 data shown previously,
so essentially none of the NLO generated cross section was rejected. 
Opening up the cuts to the 
value $\Delta \phi > 105$ degrees used by the E-706 experiment does not, 
therefore, result in any increase of the theoretical cross section. However, 
it does result in an increase in the experimental cross section. The 
net result is that the NLO calculation will not correctly reproduce the 
effects of different $\Delta \phi$ cuts because the theoretical distribution 
is too narrow. This results in a relative normalization shift when comparing 
experiments which used different cuts in $\Delta \phi$. This shift can be 
accomodated by altering the renormalization and factorization scales used 
in the theoretical calculations. Alternatively, one can restrict the 
comparison to data sets which use the same $\Delta \phi$ cut.
Similar considerations apply to cuts on other variables such as the net 
transverse momentum of the hadron pair, $p_{Tpair}$. The comparison 
with E-706 data \cite{Begel} is shown in FIG.~\ref{E706_qt} where one can 
see that the theoretical $p_{Tpair}$ distribution is significantly narrower 
than is observed in the data. Comparisons to data 
which integrate over the full $p_{Tpair}$ distribution will differ from 
comparisons to data sets which place cuts on this variable.

This situation should, in fact, come as no surprise. Both the $\Delta \phi$ 
and $p_{Tpair}$ distributions are examples which are delta functions at 
lowest order. Non-trivial contributions to these observables only start 
in next-to-leading order. In that sense, the curves shown here for these 
distributions are really leading-order only and they diverge at the endpoints 
corresponding to the third parton being soft and/or collinear ($\Delta \phi = 
180^{\circ} {\rm \ or\ } p_{Tpair}=0$). A more realistic treatment of these 
observables would require the application of soft gluon resummation 
techniques. Note, however, that there are compensating singularities at the 
endpoints of the distributions, so that a finite result is obtained after 
integrating over the distribution and the normalization of the integrated 
distribution is thus calculated to next-to-leading order.\footnote {Due to 
the use of finite width bins, the divergent behavior at the endpoint of the 
$p_{Tpair}$ distribution in FIG.~\ref{E706_qt} is not evident. The first bin 
remains finite as it contains the endpoint contribution, as well.}

From the standpoint of comparing to NLO calculations, it would be better if 
the data sets were defined only by cuts on variables whose distributions are 
well described by the calculation. In this sense, the E-706 procedure 
is to be preferred since a large portion of both the $\Delta \phi 
{\rm \ and \ } p_{Tpair}$ distributions were integrated over. 

The cuts utilized in the analysis of the E-706 data differ substantially from 
those which were used for the CCOR, NA-24, and E-711 data sets. Therefore, 
separate comparisons are required and the optimum choice of scale will differ 
between the two sets of experiments. It must be stressed that this is not an 
experimental problem, but rather the result of the fact that the NLO 
calculation does not properly describe the distributions in some of the 
variables used for making the cuts.

The E-711, E-706, and CCOR Collaborations each measured the angular 
distribution of 
the dihadron pair in the parton-parton center-of-momentum frame. None of the  
experiments observed any significant variation of this distribution with 
dihadron mass or with energy. The theoretical results for the normalized 
$\cos \theta^*$ distribution are compared to the E-711 data \cite{E711, White}
in FIG.~\ref{dcost4} for $7<M<7.5$ GeV and in FIG.~\ref{dcost3} for 
$M>7.5$ GeV. In each figure two curves are shown corresponding to the 
NLO results with scale choices of $0.85 M$ and $\sqrt{<p_T^2>}$, where 
$<p_T^2>$ 
is the average of the squared transverse momentum for the two 
observed hadrons in the event. Note that for the case of two-body kinematics 
at fixed $M$, the parton transverse momentum is $\frac{M}{2} \sin\theta^*$ 
and the parton and hadron transverse momenta are nearly the same since the 
fragmentation variable $z$ is near one. Thus, one can argue that for fixed 
$M$ and $\theta^*$ either $M$ or $\sqrt{<p_T^2>}$ is a valid choice for the 
scale. The choice of $\sqrt{<p_T^2>}$ gives a steeper 
distribution which is in better agreement with the data than is the 
result obtained with the choice of $0.85 M$ for the scales. This steepening 
occurs because at fixed $M$ as $\cos \theta^* \rightarrow 1, \sqrt{<p_T^2>} $ 
decreases. This decreasing scale causes an increase in the theoretical 
cross section. However, since the distribution is normalized to unity at 
$\cos \theta^*=0$, the end result is a steeper angular distribution.

\section{Summary and Conclusions}
\label{sec:summary}

   A next-to-leading-log Monte Carlo program has been constructed 
for symmetric dihadron production using the 
two cut-off phase space slicing formalism described in Refs.~\cite{bergmann,
HO}. This process serves as a probe of the underlying hard-scattering 
subprocesses which complements that provided by single particle production. 
For high mass pairs the relevant range of the fragmentation momentum 
fraction $z$ is comparable to that for single particle production. The 
results presented here show that the NLO QCD formalism is capable of 
giving a good description of the data for the dihadron mass and angular 
distributions. There appears to be no anomalous behavior with respect to 
either the dihadron mass or the center-of-mass energy. This is in marked 
contrast to the cases for direct photon and single hadron production at fixed 
target energies. 
This process, therefore, provides encouraging evidence that the underlying 
hard scattering is correctly described by QCD and that the problems with the 
single photon and single hadron cross sections may be ascribed to a 
combination of of effects due to an incomplete application of the theory and 
possible inconsistencies amongst the various data sets.

\begin{acknowledgments}
The author wishes to thank Brian Harris for useful discussions.
\end{acknowledgments}

%
%

\end{document}